# Visual Perception, Quantity of Information Function and the Concept of the Quantity of Information Continuous Splines

**Rushan Ziatdinov**

*Department of Industrial and Management Engineering*
*Keimyung University, 704-701 Daegu, Republic of Korea*
*E-mail: ziatdinov@kmu.ac.kr, rushanziatdinov@gmail.com*
*URL: http://www.ziatdinov-lab.com/*

**Abstract**

The geometric shapes of the outside world objects hide an undisclosed emotional, psychological, artistic, aesthetic and shape-generating potential; they may attract or cause fear as well as a variety of other emotions. This suggests that living beings with vision perceive geometric objects within an information-handling process. However, not many studies have been performed for a better understanding of visual perception from the view of information theory and mathematical modelling, but the evidence first found by Attneave (1954) suggests that the concepts and techniques of information theory may shed light on a better and deeper understanding of visual perception. The quantity of information function can theoretically explain the concentration of information on the visual contours, and, based on this, we first propose the concept of the quantity of information continuous splines for visualization of shapes from a given set of discrete data without adding any in-between points with curvature extreme. Additionally, we first discover planar curve with a constant quantity of information function and demonstrate one of the conditions when a monotonic curvature curve has a constant quantity of information function.

**Keywords:** visual perception; quantity of information; curvature continuity; spline; information loss; visual contour.

# Визуальное восприятие, функция количества информации и концепция сплайнов с непрерывной функцией количества информации

**Рушан Анурович Зиатдинов**

*Кафедра промышленной инженерии и инженерного менеджмента*
*Университет Кемён, 704-701 г. Тэгу, Южная Корея*
*E-mail: ziatdinov@kmu.ac.kr, rushanziatdinov@gmail.com*
*URL: http://www.ziatdinov-lab.com/*

**Аннотация**

Геометрические формы объектов окружающего мира скрывают в себе эмоционально-психологический, художественно-эстетический и формообразующий потенциал, могут привлекать своей красотой или вызывать чувство страха либо другие эмоции. Это наталкивает на мысль, что живые существа, обладающие зрением, могут воспринимать геометрические формы объектов в рамках процесса обработки информации. К сожалению, лишь небольшое число исследований были посвящены более глубокому пониманию визуального восприятия с позиций теории информации и математического моделирования, однако, результаты наблюдений, впервые полученные Атнивом (1954), приводят к умозаключению, что концепции и методы теории информации могут пролить свет на более фундаментальное понимание визуального восприятия. С помощью функции количества информации можно дать теоретическое обоснование концентрации информации на визуальных контурах и, с учетом этого понятия, нами впервые выдвигается концепция сплайнов с непрерывной функцией количества информации, которые могут быть использованы для визуализации форм по имеющимся дискретным данным, причем сегменты кривых между заданными точками не будут содержать экстремумов функции кривизны. Кроме того, в настоящей работе впервые найдено семейство плоских кривых с постоянной функцией количества информации, а также показано одно из условий, при которых кривые с монотонной функцией кривизны имеют постоянную функцию количества информации.

**Ключевые слова:** визуальное восприятие; количество информации; непрерывность кривизны; сплайн; потеря информации; визуальный контур.

# 1. Introduction

Technically, visual perception is the ability to interpret the surrounding environment by processing information that is contained in visible light. In 1954, Attneave noted that perception is an information-handling process. Several studies have paved the way to a better understanding of the visual perception of humans and apes, and we explain some of these below.

Attneave (1954) pointed out some of the ways in which the concepts and techniques of information theory may aid in an understanding of visual perception. Based on his empirical study, he famously suggested that information along visual contours is concentrated in regions of high magnitude of curvature as opposed to being distributed uniformly along the contour. He also argued that much of the information received by any higher organism is redundant. He claimed that this is because of an area of homogenous colour (colour includes brightness here) and a contour of homogenous direction or slope. It is further concentrated at those points on a contour at which its direction changes more rapidly (i.e. peaks of curvature). The observation of Attneave (1954) was informal but astute. His work helped to inspire interest in information-processing approaches to the study of vision. Attneave's experiments were never published, but Norman et al. (2001) have conducted a similar experiment and repeated Attneave's results.

Matsuno and Tomonaga (2007) examined the sensitivity of human animals (i.e. chimpanzees) to the negative (concave) or positive (convex) contour apex. The study adopted a two-alternative matching-to-sample procedure using two-dimensional polygons (Barenholtz et al., 2003) on the monitor as stimuli. Five chimpanzees were used in this experiment, consisting of both male and female; two of them aged five years old and the rest aged 28. The chimpanzees were required to distinguish the shape of polygons from a distracting stimulus. Based on the results, chimpanzees performed significantly better in the concave versus the convex deformation trials. Like humans, chimpanzees are more sensitive to a concave deformation of a perceived shape representation than to a convex deformation.

Unfortunately, the influence of shape topology on the visual perception has only minimally been studied. The first recognisable attempt was made by Yoshimoto and Harada (2002), who noted that five types of logarithmic distribution diagrams of curvature cause different types of impressions. The later work of Musa et al. (2015) included an experimental study of how different personality types (sanguine, melancholic, choleric, phlegmatic) are attracted by various shapes having a central symmetry, monotonic curvature or other characteristics.

There are some intersections of our research with the work of Farin and Sapidis (1989), who proposed that a curve is fair if its curvature plot consists of relatively few monotone pieces. However, recent theories and empirical research carried out by Nabiyev and Ziatdinov (2014) show that, from the view of technical aesthetics, not all curve segments with monotonically varying curvature can be considered fair (aesthetic). Therefore, aesthetic aspects of the quantity of information continuous splines will not be studied in this work.

The remainder of this paper is organised as follows. Section 2 consists of two subsections. In Section 2.1, we briefly describe the meaning of the quantity of information, identify certain conclusions and address the limitations of previous works. In Section 2.2, we perform an analysis of the quantity of information function and present the cases in which this function becomes constant. Then, in Section 3, we share ideas concerning the quantity of information continuous splines and discuss the curves that can be used for a spline generation. Section 4 provides several examples. Finally, in Section 4, we summarise our findings and discuss the scope for future work.

# 2. Information Concentration along Visual Contours

## 2.1 Preliminaries

This section briefly describes the meaning of the quantity of information and focuses on conclusions and limitations that derive from the works cited below.

Shannon (1948) showed the quantity of information is

$$u(M) = -\log(p(M)), \quad (1)$$

where *p(M)* is a probability density function.

Feldman and Singh (2005) considered the case of simple planar curves with no self-intersections. In their work, the change in tangential direction on a smooth curve follows a von Mises distribution (Von Mises, 1936) centred on a straight $\alpha = 0$:

$$p(\alpha) = A\exp(b\cos(\alpha)). \quad (2)$$

A particular point along the curve and a particular value of turning angle $\alpha$ information is measured as

$$u(\alpha) = -\log(p(\alpha)) = -\log A - b\cos(\alpha). \quad (3)$$

The quantity of information (surprisal) of a given value of curvature $\kappa$ can be written as follows:

$$u(\kappa) = -\log A' - b(\Delta s)^2 \cos(\kappa \Delta s), \quad (4)$$

and for closed contours becomes

$$u(\kappa) = -\log A' - b(\Delta s)^2 \cos(\kappa \Delta s - \frac{2\pi}{n}). \quad (5)$$

The quantity of information function defined as in (5) is a function composition $u \circ \kappa(L)$, where $\Delta s = L/n$ ( $L$ – length[1], $n$ is the number of uniformly spaced points on a closed contour).

The theoretical research of Feldman and Singh (2005) supports the following conclusions:
- Information increases with curvature;
- Negative curvature points carry greater information than equivalent positive-curvature points (not depending on the precise choice of the von Mises distribution);

## 2.2. Analysis of the Quantity of Information Function

According to (5), the surprisal in some cases can be a periodic function (Fig. 1) and has its minimum when the tangent direction turns slightly $\left(\frac{2\pi}{n}\right)$ inward. In the case of monotonic curvature curves (MC-curves), the amplitudes of (5) increase with an arc length (Fig. 2).

---

[1] Further on, we will use *s* to represent an arclenght.

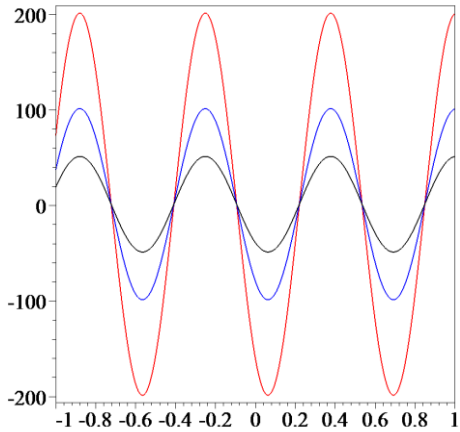

**Fig. 1.** The quantity of information function (5) with $\kappa \in [-1; 1], s = 100, n = 10, A' = 1/4$ and $b = \{1/2, 1, 2\}$ for black, blue and red lines, respectively[2].

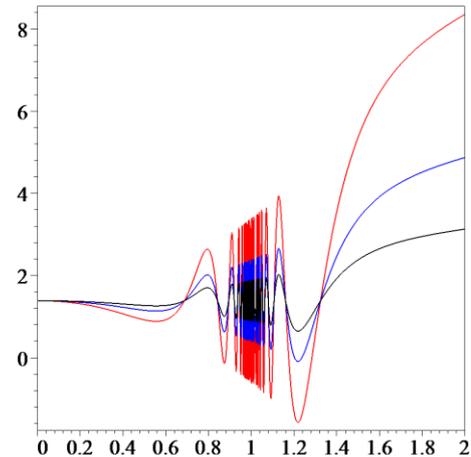

**Fig. 3.** The quantity of information function (5) for a curve with given curvature $\kappa(s) = (1 - s)^{-1}$, $s \in [0; 2], s = 100, n = 10, A' = 1/4$ and $b = \{1/2, 1, 2\}$ for black, blue and red lines, respectively.

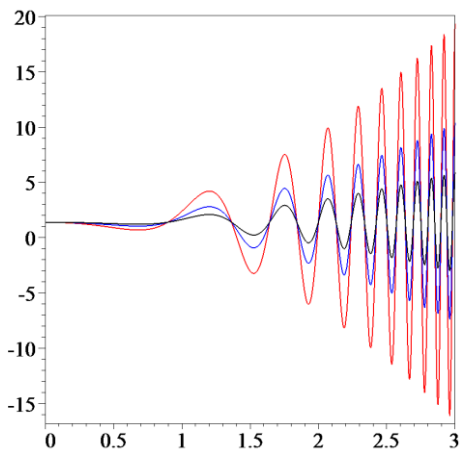

**Fig. 2.** The quantity of information function (5) for an MC-curve with given curvature $\kappa(s) = \exp(s), s \in [0; 3], s = 100, n = 10, A' = 1/4$ and $b = \{1/2, 1, 2\}$ for black, blue and red lines, respectively.

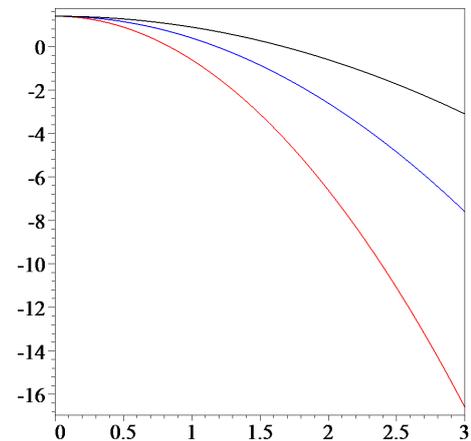

**Fig. 4.** The quantity of information function (5) for an MC-curve (clothoid) with given curvature $\kappa = 2\pi/ns$, $s \in [0; 3], s = 100, n = 10, A' = 1/4$ and $b = \{1/2, 1, 2\}$ for black, blue and red lines, respectively.

The works of Feldman and Singh (2005) as well as Attneave (1954) do not show how the information should be measured for self-intersection points, nor do they show any examples for contours containing singular points (cusps) in which $K = \infty$ (Fig. 3). Obviously, following (4), the limit of the quantity of information function becomes

$$\lim_{\kappa \to \infty} u(\kappa) = undefined. \qquad (6)$$

On the other hand, the experimental conclusions of Attneave (1954) lead to the inference that $u(\infty)$ must be a high value.

An interesting case of

$$\kappa(s) = \frac{1}{s}\left(n \arccos\left(\frac{Cn^2}{s^2}\right) + 2\pi\right), \qquad (7)$$

where $C = const, n = const$ leads to the case of $u(\kappa) \equiv const$. It is important to note that from (7) it can be determined that

$$-1 \leq \frac{Cn^2}{s^2} \leq 1, \text{ or } -s^2 \leq Cn^2 \leq s^2.$$

Figs. 5-6 show the case of a class of curves with a constant quantity of information function.

---

[2] For an interpretation of the references to colour in this figure and in subsequent figures, the reader is referred to the web version of this article.

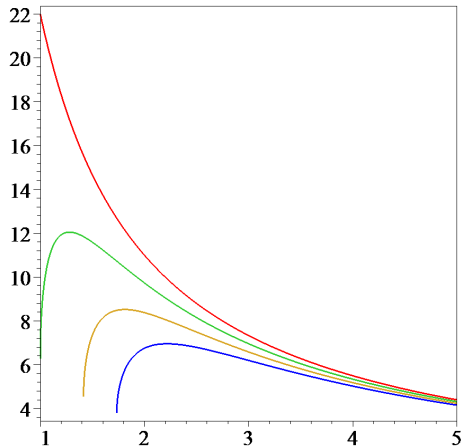

**Fig. 5.** Curvature function (7) for $C = 0, 1 \times 10^{-2}, 2 \times 10^{-2}, 3 \times 10^{-2}$ (for red, green, yellow and blue, respectively) and $n = 10, s \in [0; 5]$.

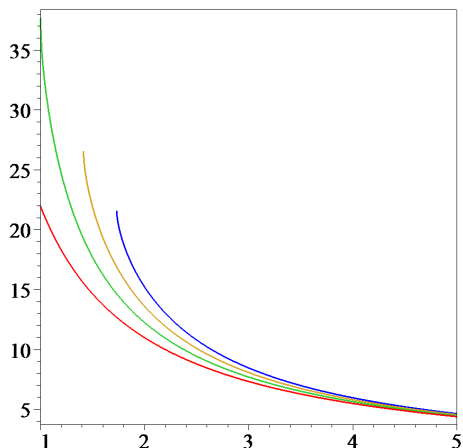

**Fig. 6.** Curvature function (7) for $C = 0, -1 \times 10^{-2}, -2 \times 10^{-2}, -3 \times 10^{-2}$ (for red, green, yellow and blue, respectively) and $n = 10, s \in [0; 5]$.

While $C \leq 0$, $\kappa(s)$ represents a product of two strictly monotonically decreasing functions, it also becomes strictly monotonic (see Fig. 6) and can be considered as a curvature function for an MC-curve.

A particular case of (7) where $C = 0 \Rightarrow \kappa(s) = \frac{\pi}{s}\left(\frac{n}{2} + 2\right)$, which inherently matches to a clothoid with $\kappa$ − shift, i.e pseudospirals with linear reparametrisation of the curvature function (Savelov, 1960), demonstrates that a spiral can be a curve with the constant quantity of information function.

## 3. The Concept of the Quantity of Information Continuous Splines

In computer graphics and computer aided geometric design, the vast majority of shapes are modelled by free-form curves and surfaces in polynomial or rational form. Usually, designers are using quadratic, cubic or quintic splines and connecting curve pieces with tangent line or curvature continuity. To prevent an *information loss* while interpolating the given set of points, we propose the concept of the quantity of information continuous splines (*QIC splines*) which should strictly satisfy the following conditions:

- *Necessary condition.* Curve segments should be connected with curvature continuity. Since the quantity of information function (5) is a composition function $u \circ$ $\kappa(s)$, it will be continuous when $\kappa(s)$ is continuous (Fikhtengolts, 1957).

- *Additional condition.* All connected segments in a QIC-spline should have a monotonically varying curvature, which means that they should be one of the majorities of MC-curves (Ziatdinov, 2012; Ziatdinov et al., 2013). This condition is necessary to prevent some in-between points in which the curvature function has its extreme values (from the work of Attneave (1954) it follows that these points may add some unnecessary information). The shape of interpolating curve segments may change, but according to our approach, it should not affect the perception of an observer in terms of being able to understand the shown objects.

The MC-curves mentioned above (Ziatdinov, 2012; Ziatdinov et al., 2013) represent many smooth curves such as pseudospirals, log-aesthetic curves (LAC), generalized log-aesthetic curves (GLAC), LAC triplets and superspirals and have applications in CAD and mechanical engineering. In this work, we consider the family of MC-curves as the first candidates for the quantity of information continuous splines, and we briefly discuss two of them.

### 3.1 Pseudospirals and Log-aesthetic Curves

Pseudospirals are planar spirals whose natural equations can be written in the form (Savelov, 1960):

$$\rho = \alpha s^m \Rightarrow \kappa = \frac{1}{\alpha} s^{-m}, (\alpha \in \mathbb{R}, m \in \mathbb{R}), \qquad (8)$$

where $\rho$ is the radius of curvature, $\kappa$ is the curvature and $s$ is the arc length. When $m = 1$, this is called the logarithmic spiral; when $m = -1$, this is referred to as the Cornu spiral; and when $= 1/2$, it is the involute of a circle. Supposing the pseudospiral has $\theta = 0$ when $s = 0$, and integrating in the well-known relationship in differential geometry, $\frac{\partial \theta}{\partial s} = \rho$ (Pogorelov, 1974), we find the tangent angle at every point of a pseudospiral.

The linear transformation of Eq. 8 leads to the natural equation of log-aesthetic curves (LACs) which are used in computer-aided design to represent a visually pleasing shape (Yoshida et al., 2006):

$$\rho(s) = \begin{cases} e^{\lambda s}, \alpha = 0 \\ (\lambda \alpha s + 1)^{\frac{1}{\alpha}}, \text{otherwise} \end{cases}. \qquad (9)$$

The analytic parametric equations of LACs were represented by incomplete Gamma functions in (Ziatdinov et al., 2012).

Since the log-aesthetic curve is not always drawable, as Yoshida et al. (2006) showed, it may fail to interpolate a set of given points with given tangent directions or curvature values. Miura et al. (2013) have proposed a new method for generating an S-shaped log-aesthetic spline and a method for solving the $G^2$ Hermite interpolation problem with log-aesthetic curves in the form of log-aesthetic triplets. These methods have been implemented as a plug-in module for a commercial CAD system and used successfully for practical design. However, due to several of the constraints of a log-aesthetic curve and the absence of precise information on tangent directions at given points, these methods have failed to interpolate the object in Fig. 10[3].

### 3.2 Superspirals

Superspirals (Ziatdinov, 2012) are a new and very wide family of fair curves, whose radius of curvature is given by a completely monotonic Gauss hypergeometric function defined in Eq. 10. The superspirals are generalisations of LACs as well as many other curves whose radius of curvature is a particular case of a completely monotonic Gauss hypergeometric function. They include a substantial variety of fair curves with monotonic curvatures and can be computed with a high degree of accuracy using the adaptive Gauss–Kronrod numerical integration method.

---

[3] Private communication with Prof. Kenjiro T. Miura and Mr. Sho Suzuki.

Ziatdinov (2012) defined a superspiral as a planar curve with a completely monotone radius of curvature given in the form $\rho(\psi) = {}_2F_1(a,b,c,-\psi)$, where $c > b > 0, a > 0$. Its corresponding parametric equations in terms of the tangent angle are

$$S(a,b,c,\theta) = \begin{pmatrix} x(\theta) \\ y(\theta) \end{pmatrix} = \begin{pmatrix} \int_0^\theta {}_2F_1(a,b,c,-\psi) \cos\psi \, d\psi \\ \int_0^\theta {}_2F_1(a,b,c,-\psi) \sin\psi \, d\psi \end{pmatrix}, \quad (10)$$

where $0 \leq \theta < +\infty$. This curve can be used as a transition curve, as it was shown in (Ziatdinov, 2012).

## 4. Examples

Figs. 7 to 9 show a car body designed using different splines. The Yoshida-Saito interpolation method (Yoshida et al., 2006) was used in the examples in Fig. 7. $G^1$ log-aesthetic splines allow the creation of high-quality shapes (Farin et al., 1989) without loops, cusps, or oscillations. Obviously, the visual perception changes with the increase in the polynomial degree, which is not likely recommended for use in aesthetic shape modelling because of the well-known Runge's phenomenon (Fornberg et al., 2007) as well as difficulty controlling curvature extremes. Figs. 8 to 9 contain such extreme points, which form unnecessary corners.

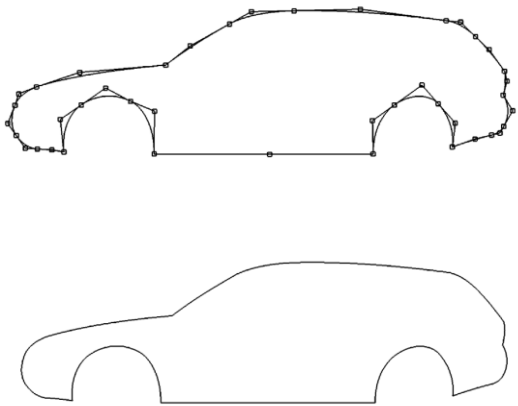

**Fig. 7.** An example of a car body design with $G^1$ log-aesthetic splines, N=42 (adapted from (Ziatdinov et al., 2012) and originally created by Prof. Norimasa Yoshida).

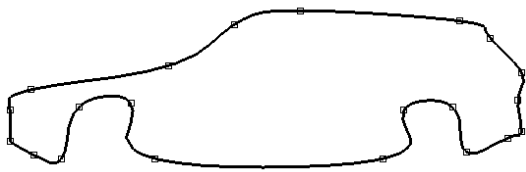

**Fig. 8.** An example of a car body design generated by a degree 2 piecewise polynomial spline, N=21.

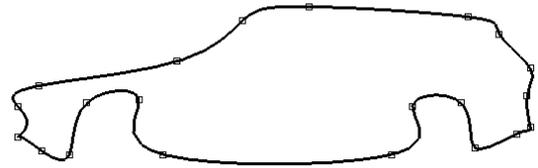

**Fig. 9.** An example of a car body design generated by a degree 3 piecewise polynomial spline, N=21.

Fig. 10 shows the shape of a duck created by a polynomial spline with not-a-knot end conditions based on N=30 given points. The curvature extremes may appear in-betweens of the given points, and this leads to the change of given information. For example, the duck's head in Fig. 10 resembles the head of a Pterosaur (Lawson, 1975).

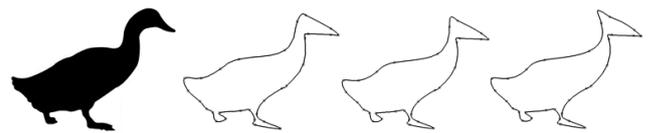

**Fig. 10.** A duck which "turns into" a Pterosaur with the increase of the polynomial degree (from left to right n=3, 4 and 5).

## 5. Conclusion and Future Work

In this work, we have:

- Analysed the quantity of information function in terms of its parameters;

- Discovered the family of curves with a constant quantity of information function;

- Found one of the conditions when an MC-curve has a constant quantity of information function;

- Proposed a new concept of the quantity of information continuous splines and presented some candidate curves for this concept;

- Shown that the visual appearance of an object might change as a result of a change in the polynomial degree.

In practice, it is not an easy task to interpolate a set of given points inclusive of their tangent directions as well as curvature values and to consider that generated curve segments will have monotonically varying curvature. Simply put, such uniform and flawless methods are not yet discovered. We believe that superspirals (Ziatdinov, 2012), which have several degrees of freedom, are one of the good candidates for such splines. Our recently proposed method of interpolation in biangular coordinates (Ziatdinov et al., 2015) can be one of the possible directions of our research on the quantity of information continuous splines.

Finally, we are very curious about one question: "Are there any objects in nature that have a contour with a constant quantity of information function?".

## 6. Acknowledgments

This research was initiated in August 2014 and I was supported by the National University Corporation – Shizuoka University, Hamamatsu, Japan for a visit to Prof. Kenjiro Miura's Realistic Modeling Laboratory. I am grateful to Prof. Kenjiro T. Miura and Mr. Sho Suzuki for providing the point coordinates for Fig. 10. Additionally, I am grateful to Prof. Sevinç Gülseçen for her kind invitation to visit the Department of Informatics at Istanbul University in Turkey and discuss important ideas related to visual perception.

The author appreciates the issues and remarks of the anonymous reviewers which helped to improve the quality of this paper.

## References

[1] Barenholtz, E., Cohen, E.H., Feldman, J., and Singh, M. (2003). Detection of change in shape: an advantage for concavities, Cognition 89, 1–9.

[2] Farin, G. and Sapidis, N. (1989). Curvature and the fairness of curves and surfaces, IEEE Computer Graphics and Applications 9 (2), 52–57.

[3] Feldman, J., and Singh, M. (2005). Information along contours and object boundaries, Psychological Review 112(1), 243–252.

[4] Fikhtengolts, G. M. (1957). Principles of mathematical analysis, State Publishing House of Technical and Theoretical Literature, Volume I, Moscow, USSR.

[5] Fornberg, B., and Zuev, J. (2007). The Runge phenomenon and spatially variable shape parameters in RBF interpolation. Computers & Mathematics with Applications, 54(3), 379-398.

[6] Lawson, D. A. (1975). Pterosaur from the latest Cretaceous of west Texas: discovery of the largest flying creature. Science, 187(4180), 947-948.

[7] Matsuno, T. and Tomonaga, M. (2007). An advantage for concavities in shape perception by chimpanzees (Pan troglodytes), Behavioral Processes 75, 253-258.

[8] Miura, K. T., Shibuya, D., Gobithaasan, R. U., and Usuki, S. (2013). Designing Log-aesthetic Splines with $G^2$ continuity. Computer-Aided Design and Applications, 10(6), 1021-1032.

[9] Miura, K.T., Gobithaasan, R.U. (2009). Generalization of Log-Aesthetic Curve, 12th International Conference on Humans and Computers, 951-952.

[10] Musa, S., Ziatdinov, R., Sozcu, O. F., and Griffiths, C. (2015). Developing Educational Computer Animation Based on Human Personality Types, European Journal of Contemporary Education 11(1), 52-71.

[11] Nabiyev, R.I. and Ziatdinov, R. (2014). A mathematical design and evaluation of Bernstein-Bézier curves' shape features using the laws of technical aesthetics, Mathematical Design & Technical Aesthetics 2(1), 6-13.

[12] Norman, J. F., Norman, F., and Ross, H. E. (2001). Information concentration along the boundary contours of naturally shaped solid objects, Perception 30, 1285–1294.

[13] Pogorelov, A. (1974). Differential Geometry, Nauka, Moscow, USSR.

[14] Regolin, L., Tommasi, L., and Vallortigara, G. (2000). Visual perception of biological motion in newly hatched chicks as revealed by an imprinting procedure, Animal Cognition 3(1), 53-60.

[15] Savelov, A. A. (1960). Planar curves, GIFML: Moscow, USSR (in Russian).

[16] Shannon, C. (1948). A mathematical theory of communication, The Bell System Technical Journal 27, 379-423.

[17] Von Mises, R. (1936). La distribution de la plus grande de *n* valeurs. Rev. Math. Union Interbalcanique 1(1).

[18] Yoshida, N., and Saito, T. (2006). Interactive aesthetic curve segments. The Visual Computer 22(9-11), 896-905.

[19] Yoshimoto, F., and Harada, T. (2002). Analysis of the characteristics of curves in natural and factory products. In Proc. of the 2nd IASTED International Conference on Visualization, Imaging and Image Processing (pp. 276-281).

[20] Ziatdinov, R. (2012). Family of superspirals with completely monotonic curvature given in terms of Gauss hypergeometric function, Computer Aided Geometric Design 29(7), 510–518.

[21] Ziatdinov, R., Kim, T. W., and Nabiyev, R. I. (2015). Two-point $G^1$ Hermite interpolation in biangular coordinates. Journal of Computational and Applied Mathematics 287, 1-11.

[22] Ziatdinov, R., Nabiyev, R.I., and Miura, K.T. (2013). MC-curves and aesthetic measurements for pseudospiral curve segments, Mathematical Design & Technical Aesthetics 1(1), 6-17.

[23] Ziatdinov, R., Nabiyev, R.I., and Miura, K.T. (2013). On some families of planar curves with monotonic curvature function, their aesthetic measures and applications in industrial design, Bulletin of Moscow Aviation Institute 20 (2), 209-218.

[24] Ziatdinov, R., Yoshida, N., and Kim, T. (2012). Analytic parametric equations of log-aesthetic curves in terms of incomplete gamma functions. Computer Aided Geometric Design 29 (2), 129–140.